\documentclass[11pt]{article}
\usepackage{amssymb,latexsym,amsmath,amsbsy,amsthm}
\usepackage[dvips]{graphicx}
\usepackage{xcolor}
\usepackage{cite}
\usepackage{stmaryrd}  
\usepackage{arydshln}  
\usepackage{mathdots}  
\def\nn{\nonumber}
\def\deg{\mathop{\rm deg}\nolimits}

\def\spn{\mathop{\rm span}\nolimits}
\def\qdots{\mathinner{\mkern1mu\raise1pt\vbox{\kern7pt\hbox{.}}\mkern2mu \raise4pt\hbox{.}\mkern2mu\raise7pt\hbox{.}\mkern1mu}}
\def\Z{{\mathbb Z}}

\def\ssl{\mathfrak{sl}}

\def\g{\mathfrak{g}}

\def\so{\mathfrak{so}}

\def\sp{\mathfrak{sp}}

\def\lb{\llbracket}
\def\rb{\rrbracket}

\def\beq{\begin{equation}}
\def\eeq{\end{equation}}

\begin{document}
\noindent
{\Large \bf
On classical $\Z_2\times\Z_2$-graded Lie algebras} \\[5mm]
{\bf N.I.~Stoilova$^{1,a}$ and J.\ Van der Jeugt$^{2,b}$} \\[2mm]
$^1$Institute for Nuclear Research and Nuclear Energy, Bulgarian Academy of Sciencies,\\ 
Boul.\ Tsarigradsko Chaussee 72, 1784 Sofia, Bulgaria\\
$^2$Department of Applied Mathematics, Computer Science and Statistics, Ghent University,\\
Krijgslaan 281-S9, B-9000 Gent, Belgium\\[2mm]
{ $^{a)}$
stoilova@inrne.bas.bg\\
{ $^{b)}$
Joris.VanderJeugt@UGent.be}

\vskip 2 cm

\noindent
{\bf ABSTRACT}

\vskip 0.5cm
\noindent 
We construct classes of $\Z_2\times\Z_2$-graded Lie algebras corresponding to the classical Lie algebras, in terms of their defining matrices.
For the $\Z_2\times\Z_2$-graded Lie algebra of type $A$, the construction coincides with the previously known class.
For the $\Z_2\times\Z_2$-graded Lie algebra of type $B$, $C$ and $D$ our construction is new and gives rise to interesting defining matrices closely related to the classical ones but undoubtedly different.
We also give some examples and possible applications to parastatistics.

\vskip 10mm


\section{INTRODUCTION} \label{sec:Introduction}%

Colour algebras and colour superalgebras are generalizations of Lie algebras and Lie superalgebras, 
introduced by Rittenberg and Wyler~\cite{Rit1, Rit2}.
For such structures, the algebra is graded by some abelian grading group $\Gamma$.
The simplest case not coinciding with a Lie superalgebra is for $\Gamma=\Z_2\times\Z_2$.
For an algebra graded by $\Z_2\times\Z_2$, there are already two distinct choices for the Lie bracket.
These correspond to structures that are nowadays known as $\Z_2\times\Z_2$-graded Lie algebras on the one hand, 
and $\Z_2\times\Z_2$-graded Lie superalgebras on the other hand.
This terminology is slightly misleading, since these algebras are not Lie algebras nor Lie superalgebras.
But since these terms have become common now in literature, we shall also stick to them.

Compared to the central role of Lie algebras and Lie superalgebras, 
the $\Z_2\times\Z_2$-graded Lie (super)algebras received for many years little attention in theoretical 
and mathematical physics~\cite{LR1978,Vasiliev1985,JYW1987}.
Only in recent years there is renewed interest in $\Z_2\times\Z_2$-graded Lie (super)algebras.
For example, such structures have appeared in 
symmetries of L\'evy–Leblond equations~\cite{Aizawa1,Aizawa3},
in graded (quantum) mechanics and quantization~\cite{Bruce2,AMD2020,Aizawa4,Aizawa5,Quesne2021},
and in $\Z_2\times\Z_2$-graded two-dimensional models~\cite{Bruce1,Toppan1,Bruce3}.
In particular $\Z_2\times\Z_2$-graded Lie algebras and superalgebras have been recognized in 
parastatistics~\cite{Tolstoy2014,SV2018} and in the description of parabosons and parafermions~\cite{Toppan2,Toppan3}.

One of the reasons why $\Z_2\times\Z_2$-graded Lie (super)algebras are considered to be more fundamental than more general colour (super)algebras, is because they still provide algebras with only commutators and anticommutators, just like Lie algebras and Lie superalgebras.
Bearing in mind that commutators and anticommutators are the essential brackets for two physical operators $x$ and $y$, it is reasonable to stick to algebraic structures with such bracket relations.
But there is a reason to go beyond Lie algebras and Lie superalgebras.
Indeed, if one considers two elements $x$ and $y$ of an associative algebra 
(out of which Lie algebras or Lie superalgebras are usually constructed), then there are two ways to rewrite the trivial product identity in this algebra in terms of commutators and anticommutators:
\[
[x,y]+[y,x]=0,\qquad \{x,y\}-\{y,x\}=0.
\]
But if one considers three elements $x$, $y$ and $z$ of an associative algebra then there are essentially 
(i.e.\ up to permuting elements) four ways~\cite{YJ} to rewrite the trivial product identity among these elements:
\begin{align*}
& [x,[y,z]]+[y,[z,x]]+[z,[x,y]]=0, \\
& [x,\{y,z\}]+[y,\{z,x\}]+[z,\{x,y\}]=0, \\
& [x,\{y,z\}]+\{y,[z,x]\}-\{z,[x,y]\}=0, \\
& [x,[y,z]]+\{y,\{z,x\}\}-\{z,\{x,y\}\}=0. 
\end{align*}
The first corresponds to the Jacobi identity for Lie algebras.
The second and third appear in the Jacobi identity for Lie superalgebras, $\Z_2$-graded, with $x$ respectively odd or even.
The fourth expression can appear only in the Jacobi identity~\eqref{jacobi} for $\Z_2\times\Z_2$-graded Lie algebras 
(or for $\Z_2\times\Z_2$-graded Lie superalgebras).
This argument adds to the value of $\Z_2\times\Z_2$-graded Lie (super)algebras as being essential mathematical structures for models in mathematical physics.

In this paper we will investigate the ways in which classical Lie algebras can be extended to $\Z_2\times\Z_2$-graded Lie algebras.
Of course, certain examples or certain classes of $\Z_2\times\Z_2$-graded Lie algebras have already been considered in the literature.
For instance, Rittenberg and Wyler give a general method~\cite{Rit2} to extend any Lie algebra of dimension $D$ to a $\Z_2\times\Z_2$-graded Lie algebra of dimension $4D$. 
Our approach is different, and considers the following question: given a classical Lie algebra $G$ of dimension $D$, is there a corresponding $\Z_2\times\Z_2$-graded Lie algebra $\g$ of dimension $D$? Moreover, we want to understand in how many ways such a possible correspondence can be obtained.

In the following section we recall the definition of $\Z_2\times\Z_2$-graded Lie algebras and make some general considerations. 
In Section~\ref{sec:C} we explain our method of construction, and obtain $\Z_2\times\Z_2$-graded Lie algebra analogues of the classical Lie algebras of type $A$, $B$, $C$ and $D$, i.e.\ of $\ssl(n+1)$, $\so(2n+1)$, $\sp(2n)$ and $\so(2n)$.
For the case of $\ssl(n+1)$ the corresponding $\Z_2\times\Z_2$-graded Lie algebra is not new, and has already appeared as a sequence in~\cite{Rit2}.
The $\Z_2\times\Z_2$-graded Lie algebras corresponding to the classical Lie algebras $\so(2n+1)$, $\sp(2n)$ and $\so(2n)$ are new, as far as we know.
Some examples and special cases of these $\Z_2\times\Z_2$-graded Lie algebras are given in Section~\ref{sec:D}. These examples are inspired by physics, and in particular by parastatistics relations.

Let us finally mention that a lot is known about gradings of Lie algebras and of matrix algebras, 
and the literature on these topics in pure mathematical journals is extensive (\cite{Elduque,Bahturin,Patera} and references therein).
Many of these works emphasize on so-called fine gradings.
Moreover, the subject is always gradings of Lie algebras.
The ``$\Z_2\times\Z_2$-graded Lie algebras'' of this paper do not appear there, 
because they are not Lie algebras but graded algebras with a bracket satisfying different properties (see~\eqref{symmetry}-\eqref{jacobi}).

\section{$\Z_2\times\Z_2$-GRADED LIE ALGEBRAS}

The definition of  $\Z_2\times\Z_2$-graded Lie algebras (LA's) as well as examples of such algebras 
have already been given in~\cite{Rit1,Rit2}.

The $\Z_2\times\Z_2$-graded LA $\g$, as a linear space, is a direct sum of four subspaces:
\begin{equation}
\g=\bigoplus_{\boldsymbol{a}} \g_{\boldsymbol{a}} =
\g_{(0,0)} \oplus \g_{(0,1)} \oplus \g_{(1,0)} \oplus \g_{(1,1)} 
\label{ZZgrading}
\end{equation}
where $\boldsymbol{a}=(a_1,a_2)$ is an element of $\Z_2\times\Z_2$.
Elements of $\g_{\boldsymbol{a}}$ are denoted by $x_{\boldsymbol{a}}, y_{\boldsymbol{a}},\ldots$,
and $\boldsymbol{a}$ is called the degree, $\deg x_{\boldsymbol{a}}$, of $x_{\boldsymbol{a}}$.
Such elements are called homogeneous elements.
The $\Z_2\times\Z_2$-graded LA $\g$ admits a bilinear operation  $\lb\cdot,\cdot\rb$ which
satisfies the grading, symmetry and Jacobi identities:
\begin{align}
& \lb x_{\boldsymbol{a}}, y_{\boldsymbol{b}} \rb \in \g_{\boldsymbol{a}+\boldsymbol{b}}, \label{grading}\\
& \lb x_{\boldsymbol{a}}, y_{\boldsymbol{b}} \rb = -(-1)^{\boldsymbol{a}\cdot\boldsymbol{b}} 
\lb y_{\boldsymbol{b}}, x_{\boldsymbol{a}} \rb, \label{symmetry}\\
& \lb x_{\boldsymbol{a}}, \lb y_{\boldsymbol{b}}, z_{\boldsymbol{c}}\rb \rb =
\lb \lb x_{\boldsymbol{a}}, y_{\boldsymbol{b}}\rb , z_{\boldsymbol{c}} \rb +
(-1)^{\boldsymbol{a}\cdot\boldsymbol{b}} \lb y_{\boldsymbol{b}}, \lb x_{\boldsymbol{a}}, z_{\boldsymbol{c}}\rb \rb,
\label{jacobi}
\end{align} 
where
\begin{equation}
\boldsymbol{a}+\boldsymbol{b}=(a_1+b_1,a_2+b_2)\in \Z_2\times\Z_2, \qquad
\boldsymbol{a}\cdot\boldsymbol{b} = a_1b_2-a_2b_1.
\label{sign}
\end{equation}
Note that another choice for $\boldsymbol{a}\cdot\boldsymbol{b}$, namely $\boldsymbol{a}\cdot\boldsymbol{b} = a_1b_1+a_2b_2$, 
would yield the definition of a $\Z_2\times\Z_2$-graded Lie superalgebra.
We emphasize that in this paper we are only dealing with $\Z_2\times\Z_2$-graded LA's, 
hence we are always using~\eqref{sign}.

By~\eqref{symmetry}, the bilinear operation  $\lb\cdot,\cdot\rb$ for homogeneous elements is either a commutator or an anticommutator.
In addition $\g_{(0,0)}$ is a Lie subalgebra of the 
$\Z_2\times\Z_2$-graded Lie algebra $\g$ and
$ \g_{(0,1)}$,  $\g_{(1,0)}$ and $\g_{(1,1)}$ are $\g_{(0,0)}$-modules.

It will be useful to list the brackets among the subspaces more explicitly:
\begin{equation}
[\g_{(0,0)}, \g_{\boldsymbol{a}}]\subset \g_{\boldsymbol{a}}, \quad 
[\g_{\boldsymbol{a}}, \g_{\boldsymbol{a}}]\subset \g_{(0,0)}, \qquad \boldsymbol{a}\in\Z_2\times\Z_2
\end{equation}
and
\begin{equation}
\{\g_{\boldsymbol{a}}, \g_{\boldsymbol{b}}\}\subset \g_{\boldsymbol{c}}
\end{equation}
if $\boldsymbol{a}$, $\boldsymbol{b}$ and $\boldsymbol{c}$ are mutually distinct elements of $\{ (1,0),(0,1),(1,1)\}$.
As a consequence, if $\g=\g_{(0,0)} \oplus \g_{(0,1)} \oplus \g_{(1,0)} \oplus \g_{(1,1)}$ is a $\Z_2\times\Z_2$-graded LA, then any permutation of the last three subspaces maps $\g$ into another $\Z_2\times\Z_2$-graded LA.
We will refer to such mappings as ``trivial permutation transformations'' of $\g$.
 
In this paper, it will be natural to impose the condition that $\g$ is generated by 
$\g_{(1,0)}\oplus \g_{(0,1)}$. Then one can deduce from the Jacobi identity that
\begin{equation}
\g_{(0,0)} = \lb\g_{(1,0)},\g_{(1,0)}\rb + \lb\g_{(0,1)},\g_{(0,1)}\rb \qquad\hbox{and}\qquad
\g_{(1,1)} = \lb\g_{(1,0)},\g_{(0,1)}\rb.
\label{g10g01}
\end{equation}
Of course, we exclude the trivial cases where $\g_{(1,0)}$ or $\g_{(0,1)}$ are zero.

Let $\g$ be an associative algebra with a product denoted by $x\cdot y$, and suppose $\g$ has a $\Z_2\times\Z_2$-grading of the form~\eqref{ZZgrading} that is compatible with the product, i.e.\ $x_{\boldsymbol{a}} \cdot y_{\boldsymbol{b}} \in \g_{\boldsymbol{a}+\boldsymbol{b}}$.
Then it is an easy exercise to verify that the following bracket turns $\g$ into a $\Z_2\times\Z_2$-graded LA:
\begin{equation}
\lb x_{\boldsymbol{a}} , y_{\boldsymbol{b}}\rb = x_{\boldsymbol{a}} \cdot y_{\boldsymbol{b}}- 
(-1)^{\boldsymbol{a}\cdot\boldsymbol{b}} y_{\boldsymbol{b}} \cdot x_{\boldsymbol{a}}\ ,
\label{ZZbracket}
\end{equation}
i.e.\ for a bracket derived from an associative product the Jacobi identity~\eqref{jacobi} is automatically satisfied.

\section{CONSTRUCTION OF CLASSICAL  $\Z_2\times\Z_2$-GRADED LIE ALGEBRAS} \label{sec:C}%

The purpose of this paper is to investigate $\Z_2\times\Z_2$-graded LA's associated to the classical LA's $\ssl(n+1)$, $\so(2n+1)$, $\sp(2n)$ and $\so(2n)$.
Classical LA's are usually specified in terms of their defining matrix representation, 
from which the LA bracket is determined~\cite{Humphreys,Fulton}.
For $\ssl(n+1)$ the defining matrices are just the traceless $(n+1)\times (n+1)$ matrices. For the other classical LA's, one uses:
\begin{equation}
\begin{array}{llll}
G=\so(2n+1) & \begin{array}{c c}
    \begin{array} {@{} c c c @{}} \ \ n\  & \ \ n\  & 1\ \end{array} & {} \\  
    \left(
       \begin{array}{@{} c  c  c @{}}
        a & b & c\\ 
        d & -a^t & e\\
				-e^t & -c^t & 0
       \end{array}
    \right)  & \hspace{-1em}
		\begin{array}{c}
     n \\  n \\	1 
    \end{array} \\ 
    \mbox{} 
  \end{array} 
& \hbox{$b$ and $d$ antisymmetric} & (\dim G=2n^2+n); \\
G=\sp(2n) & \begin{array}{c c}
 \begin{array} {@{} c c @{}} \, n\  & \, n\  \end{array} & {} \\  
    \left(
       \begin{array}{@{} c   c @{}}
        a & b \\ 
        c & -a^t
       \end{array}
    \right)  & \hspace{-1em}
		\begin{array}{c}
     n \\  n 
    \end{array} \\ 
    \mbox{} 
  \end{array} 
& \hbox{$b$ and $c$ symmetric} &(\dim G=2n^2+n);\\
G=\so(2n) & \begin{array}{c c}
    \begin{array} {@{} c c @{}} \, n\  & \, n\  \end{array} & {} \\  
    \left(
       \begin{array}{@{} c   c @{}}
        a & b \\ 
        c & -a^t
       \end{array}
    \right)  & \hspace{-1em}
		\begin{array}{c}
     n \\  n 
    \end{array} \\ 
    \mbox{} 
  \end{array} 
&\hbox{$b$ and $c$ antisymmetric} &(\dim G=2n^2-n),
\end{array}
\label{G}
\end{equation}
where $a^t$ is the transpose of $a$.
The size of the matrices in these expressions is in an obvious way determined by the borders.

Note that for orthogonal Lie algebras $\so(N)$ (where $N$ can be even or odd) the defining matrices are often chosen as the set of antisymmetric $N\times N$ matrices.
This is certainly a valid choice, and the matrix forms are even simpler than in~\eqref{G}.
However, the advantage~\cite[p.~269]{Fulton} of~\eqref{G} is that the Cartan subalgebra consists of all diagonal matrices in the form~\eqref{G}.
In the set of antisymmetric matrices, however, there are no nonzero diagonal matrices.
Also for identification of Lie algebra elements with creation and annihilation operators of parastatistics, it is necessary to work with defining matrices of the form~\eqref{G}.

How can one implement a $\Z_2\times\Z_2$-grading on these classical LA's such that the algebra closes under the bracket~\eqref{ZZbracket}?
It is not simply a question of finding all $\Z_2\times\Z_2$-gradings of the classical LA compatible with the LA bracket, since under the new $\Z_2\times\Z_2$-graded LA bracket~\eqref{ZZbracket} the resulting matrices might no longer be of the proper form.

One can, however, start from a set of generators of the classical LA (in the defining matrix form), associate a $\Z_2\times\Z_2$-grading on these generators, compute new elements with these generators using the bracket~\eqref{ZZbracket}, and see which matrix structures and algebras arise in this way.

In order to investigate this systematically, let us assume (without loss of generality) that the generating subspace $A$ of the classical LA $G$ corresponds to the subspace $\g_{(1,0)}\oplus \g_{(0,1)}$ of the associated $\Z_2\times\Z_2$-graded LA $\g$, and that $\g$ is generated by this subspace according to~\eqref{g10g01}.
Thus we are looking for generating subspaces $A$ of a classical LA $G$ such that $G=A + [A,A]$ (as vector space).
Such generating subspaces were found in~\cite{SV2005LA}. In that paper, all so-called 5-gradings $G_{-2}\oplus G_{-1}\oplus G_0\oplus G_1\oplus G_2$ of $G$ were determined such that $G$ is generated by $A=G_{-1}\oplus G_1$ (subject to a conjugacy condition).
Following these natural assumptions, one can now proceed as follows, starting from the classical LA $G$ in its defining matrix representation:
\begin{itemize}
\item For each of the 5-gradings of $G$, let $A=G_{-1}\oplus G_1$ (as a subspace of the vector space of $G$).
\item Partition $A$ in all possible ways in two subspaces $\g_{(1,0)}\oplus \g_{(0,1)}$.
\item Construct from here the matrix elements of the $\Z_2\times\Z_2$-graded LA $\g$ using the bracket~\eqref{ZZbracket} and~\eqref{g10g01}.
\end{itemize}
This construction process is straightforward but very elaborate. 
It is clear that it will yield the matrix form of some $\Z_2\times\Z_2$-graded LA $\g$. 
But we should also admit that the list of $\Z_2\times\Z_2$-graded LA's thus obtained might not be exhaustive.

Rather than giving all details of the calculations, we restrict ourselves to listing here the final outcome.

\vskip 6mm
When $G=\ssl(n+1)$, there are 7 types of 5-gradings to be investigated~\cite{SV2005LA}. 
All cases lead to the following type of $\Z_2\times\Z_2$-graded LA (or special cases thereof). 
Let $p,q,r,s$ be integers with $p+q+r+s=n+1$. 
The $\Z_2\times\Z_2$-graded LA $\g=\ssl_{p,q,r,s}(n+1)$ consists of all traceless $(n+1)\times (n+1)$ matrices of the following block form:
\begin{equation}
\begin{array}{c c}
    \begin{array} {@{} c c  cc @{}} \ \ p\ \ & \ \ q\ \ & \ \ r\ \ & \ \ s \ \ \end{array} & {} \\  
\left(\begin{array}{cccc} 
a_{(0,0)} & a_{(0,1)} & a_{(1,0)} & a_{(1,1)} \\ 
b_{(0,1)} & b_{(0,0)} & b_{(1,1)} & b_{(1,0)} \\ 
c_{(1,0)} & c_{(1,1)} & c_{(0,0)} & c_{(0,1)} \\ 
d_{(1,1)} & d_{(1,0)} & d_{(0,1)} & d_{(0,0)} 
\end{array}\right)
 & \hspace{-1em}
		\begin{array}{l}
     p \\  q \\	r \\ s
    \end{array} \\ 
    \mbox{} 
  \end{array} \\[-12pt]  . 
\label{ZZsl}
\end{equation}	
The indices of the matrix blocks refer to the $\Z_2\times\Z_2$-grading. 
The dimensions of the four subspaces of $\g=\ssl_{p,q,r,s}(n+1)$ are given by
\[
\begin{array}{ll}
\g_{(0,0)} & p^2+q^2+r^2+s^2-1\\
\g_{(0,1)} & 2pq+2rs\\ 
\g_{(1,0)} & 2pr+2qs\\
\g_{(1,1)} & 2qr+2ps
\end{array}
\]
leading to a total dimension of $n^2+2n$. 
The matrices of $\g=\ssl_{p,q,r,s}(n+1)$ are the same as the matrices of $G=\ssl(n+1)$; this will not be the case for the $\Z_2\times\Z_2$-graded LA's associated with the other classical LA's.
All the cases from our procedure lead to~\eqref{ZZsl} or trivial permutation transformations thereof.

For the current case, one can say something more. 
Indeed, the underlying vector space is just the space of traceless matrices.
But for this matrix algebra, all $\Z_2\times\Z_2$-gradings (as an associative algebra) have been classified~\cite[Example~2.30]{Elduque}, corresponding exactly to the gradings~\eqref{ZZsl}.
Then, using the bracket~\eqref{ZZbracket} (checking closure in the same vector space, and yielding the complete vector space) one obtains also in this way the $\Z_2\times\Z_2$-graded LA's $\ssl_{p,q,r,s}(n+1)$.

This class of $\Z_2\times\Z_2$-graded LA's is not new, and has been known for a long time. 
It appears already in~\cite{Rit2} under the name $sl(p,q,r,s)$.

In the same paper~\cite{Rit2}, one can deduce the matrix form of another class of $\Z_2\times\Z_2$-graded LA's, $so(p,q,r,s)$, with $p+q+r+s=N$ 
(although the definition is not explicitly in this reference). 
The $\Z_2\times\Z_2$-graded LA's $so(p,q,r,s)$ consists of all $N\times N$ matrices of the following form:
\begin{equation}
\begin{array}{c c}
    \begin{array} {@{} c c  cc @{}} \ p\ \ & \ \ \ q\ \ & \ \ \ \ r\ \ & \ \ \ \ s \ \ \end{array} & {} \\  
\left(\begin{array}{cccc} 
a_{(0,0)} & a_{(0,1)} & a_{(1,0)} & a_{(1,1)} \\ 
-a^t_{(0,1)} & b_{(0,0)} & b_{(1,1)} & b_{(1,0)} \\ 
-a^t_{(1,0)} & b^t_{(1,1)} & c_{(0,0)} & c_{(0,1)} \\ 
-a^t_{(1,1)} & b^t_{(1,0)} & c^t_{(0,1)} & d_{(0,0)} 
\end{array}\right)
 & \hspace{-1em}
		\begin{array}{l}
     p \\  q \\	r \\ s
    \end{array} \\ 
    \mbox{} 
  \end{array} \\[-12pt]  , 
\label{ZZso}
\end{equation}	
where $a_{(0,0)}$, $b_{(0,0)}$, $c_{(0,0)}$ and $d_{(0,0)}$ are antisymmetric matrices.
The dimensions of the four subspaces of $\g=so(p,q,r,s)$ are given by
\[
\begin{array}{ll}
\g_{(0,0)} & \frac12(p(p-1)+q(q-1)+r(r-1)+s(s-1))\\
\g_{(0,1)} & pq+rs\\ 
\g_{(1,0)} & pr+qs\\
\g_{(1,1)} & qr+ps
\end{array}
\]
Just as for the antisymmetric matrix form of orthogonal LA's, the current matrices~\eqref{ZZso} have zero diagonals, and the Cartan subalgebra cannot be identified with diagonal matrices.
In order to overcome this disadvantage and obtain $\Z_2\times\Z_2$-graded LA's associated with the matrix forms~\eqref{G}, 
we continue the method outlined in the beginning of the paragraph on the LA's of~\eqref{G}.

\vskip 6mm
When $G=\so(2n+1)$, there are 3 types of 5-gradings to be investigated~\cite{SV2005LA}. 
All cases lead to the following type of $\Z_2\times\Z_2$-graded LA, where $1\leq p < n$: 
the $\Z_2\times\Z_2$-graded LA $\g=\so_{p}(2n+1)$ consists of all matrices of the following block form:
\begin{equation}
\begin{array}{c c}
    \begin{array} {@{} c c c c c @{}} \ \ p\ \ & \ \ \ n-p\ \ & \ \ p \ \ \ & \ \ n-p\ \ & \ 1 \ \ \end{array} & {} \\
\left(\begin{array}{@{} cc:cc:c @{}} a_{(0,0)}&a_{(1,1)}&b_{(0,0)}&b_{(1,1)}&c_{(0,1)}  \\
\tilde{a}_{(1,1)}&\tilde{a}_{(0,0)}&b_{(1,1)}^{\;\;t}&\tilde{b}_{(0,0)}&c_{(1,0)}  \\\hdashline
d_{(0,0)}&d_{(1,1)}&-a_{(0,0)}^{\;\;t}&\tilde{a}_{(1,1)}^{\;\;t}&e_{(0,1)}  \\
d_{(1,1)}^{\;\;t}&\tilde{d}_{(0,0)}&a_{(1,1)}^{\;\;t}&-\tilde{a}_{(0,0)}^{\;\;t}&e_{(1,0)}  \\\hdashline
-e_{(0,1)}^{\;\;t}&-e_{(1,0)}^{\;\;t}&-c_{(0,1)}^{\;\;t}&-c_{(1,0)}^{\;\;t}&0  
\end{array}\right)
 & \hspace{-1em}
		\begin{array}{c}
     p \\ n-p \\[1mm] p \\ n-p \\[1mm] 1 
		\end{array} \\ 
    \mbox{} 
  \end{array} \\[-12pt] , 
\label{ZZso2n+1}
\end{equation}
where $b_{(0,0)}$, $\tilde{b}_{(0,0)}$, $d_{(0,0)}$ and $\tilde{d}_{(0,0)}$ are antisymmetric matrices.
As before, the indices of the matrix blocks refer to the $\Z_2\times\Z_2$-grading. 
Hence the dimensions of the four subspaces of $\g=\so_{p}(2n+1)$ are given by
\[
\begin{array}{ll}
\g_{(0,0)} & 2n^2-n-4p(n-p)\\
\g_{(0,1)} & 2p\\
\g_{(1,0)} & 2(n-p)\\ 
\g_{(1,1)} & 4p(n-p) .
\end{array}
\]
Note that the matrices of $\g=\so_{p}(2n+1)$ are not the same as the matrices of $G=\so(2n+1)$ in~\eqref{G}, even though the
total dimension of the matrix space is the same.
Of course, one can perform trivial permutation transformations to~\eqref{ZZso2n+1}.
Furthermore, it is worth noticing that some other case in our procedure leads to matrix forms which are different, such as
\begin{equation}
\begin{array}{c c}
    \begin{array} {@{} c c c c c @{}}\ \ \ p\ \ & \ \ n-p\ \ & \ \ p \ \ \ & \ \ n-p\ \ & \ 1 \ \ \end{array} & {} \\
\left(\begin{array}{@{} cc:cc:c @{}} 
a_{(0,0)}&a_{(1,1)}&b_{(0,0)}&b_{(1,1)}&c_{(0,1)}  \\
\tilde{a}_{(1,1)}&\tilde{a}_{(0,0)}&-b_{(1,1)}^{\;\;t}&\tilde{b}_{(0,0)}&c_{(1,0)}  \\\hdashline
d_{(0,0)}&d_{(1,1)}&-a_{(0,0)}^{\;\;t}&-\tilde{a}_{(1,1)}^{\;\;t}&e_{(0,1)}  \\
-d_{(1,1)}^{\;\;t}&\tilde{d}_{(0,0)}&-a_{(1,1)}^{\;\;t}&-\tilde{a}_{(0,0)}^{\;\;t}&e_{(1,0)}  \\\hdashline
-e_{(0,1)}^{\;\;t}&e_{(1,0)}^{\;\;t}&-c_{(0,1)}^{\;\;t}&c_{(1,0)}^{\;\;t}&0  
\end{array}\right)
 & \hspace{-1em}
		\begin{array}{c}
     p \\ n-p \\[1mm] p \\ n-p \\[1mm] 1 
		\end{array} \\ 
    \mbox{} 
  \end{array} \\[-12pt] , 
\label{ZZso2n+1b}
\end{equation}
where again $b_{(0,0)}$, $\tilde{b}_{(0,0)}$, $d_{(0,0)}$ and $\tilde{d}_{(0,0)}$ are antisymmetric matrices.
Notice the structural difference between~\eqref{ZZso2n+1} and~\eqref{ZZso2n+1b}.
At first sight~\eqref{ZZso2n+1b} is another $\Z_2\times\Z_2$-graded LA associated to the LA $\so(2n+1)$.
However, one can map a basis of~\eqref{ZZso2n+1b} to a basis of~\eqref{ZZso2n+1} such that the basis elements satisfy the same bracket relations. Hence, the two matrix algebras are isomorphic as $\Z_2\times\Z_2$-graded LA's.

\vskip 6mm
When $G=\sp(2n)$, there are 4 types of 5-gradings to be investigated. 
Quite surprisingly, for 3 of these 4 types, the algebra generated by the subspace $\g_{(1,0)}\oplus \g_{(0,1)}$ leads to a $\Z_2\times\Z_2$-graded LA of the form~\eqref{ZZso2n},
thus associated with $\so(2n)$ rather than $\sp(2n)$. 
The other case leads to the following type of $\Z_2\times\Z_2$-graded LA, where $1\leq p < n$: 
the $\Z_2\times\Z_2$-graded LA $\g=\sp_{p}(2n)$ consists of all matrices of the following block form:
\begin{equation}
\begin{array}{c c}
    \begin{array} {@{} c c c c @{}} \ \ \ \ p\ \ & \ \ n-p\ \ & \ p \ \ \ & \ \ n-p\ \ \end{array} & {} \\
\left(\begin{array}{@{} cc:cc @{}} a_{(0,0)}&a_{(1,0)}&b_{(1,1)}&b_{(0,1)}  \\
\tilde{a}_{(1,0)}&\tilde{a}_{(0,0)}&-b_{(0,1)}^{\;\;t}&\tilde{b}_{(1,1)}   \\\hdashline
c_{(1,1)}&c_{(0,1)}&-a_{(0,0)}^{\;\;t}&-\tilde{a}_{(1,0)}^{\;\;t}   \\
-c_{(0,1)}^{\;\;t}&\tilde{c}_{(1,1)}&-a_{(1,0)}^{\;\;t}&-\tilde{a}_{(0,0)}^{\;\;t} 
\end{array}\right)
 & \hspace{-1em}
		\begin{array}{c}
     p \\ n-p \\[1mm] p \\ n-p  
		\end{array} \\ 
    \mbox{} 
  \end{array} \\[-12pt] , 
\label{ZZsp2n}
\end{equation}
where $b_{(1,1)}$, $\tilde{b}_{(1,1)}$, $c_{(1,1)}$ and $\tilde{c}_{(1,1)}$ are symmetric matrices.
The dimensions of the four subspaces of $\g=\sp_{p}(2n)$ are given by
\[
\begin{array}{ll}
\g_{(0,0)} & p^2+(n-p)^2\\
\g_{(0,1)} & 2p(n-p)\\
\g_{(1,0)} & 2p(n-p)\\ 
\g_{(1,1)} & p(p+1)+(n-p)(n-p+1) .
\end{array}
\]
Also here, the matrices of $\g=\sp_{p}(2n)$ do not coincide with the matrices of $G=\sp(2n)$ in~\eqref{G}, 
but the dimension of the matrix spaces is the same.

\vskip 6mm
Finally, when $G=\so(2n)$, there are 5 types of 5-gradings to be investigated. 
For 3 of these 5 types, the algebra generated by the subspace $\g_{(1,0)}\oplus \g_{(0,1)}$ leads to a $\Z_2\times\Z_2$-graded LA of the form~\eqref{ZZsp2n}. 
The other two cases lead to the following type of $\Z_2\times\Z_2$-graded LA, where $1\leq p < n$: 
the $\Z_2\times\Z_2$-graded LA $\g=\so_{p}(2n)$ consists of all matrices of the following block form:
\begin{equation}
\begin{array}{c c}
    \begin{array} {@{} c c c c @{}} \ \ \ \ p\ \ & \ n-p\ \ & \ p \ \ \ & \ \ n-p\ \ \end{array} & {} \\
\left(\begin{array}{@{} cc:cc @{}} 
a_{(0,0)}&a_{(1,0)}&b_{(1,1)}&b_{(0,1)}  \\
\tilde{a}_{(1,0)}&\tilde{a}_{(0,0)}& b_{(0,1)}^{\;\;t}&\tilde{b}_{(1,1)}   \\\hdashline
c_{(1,1)}&c_{(0,1)}&-a_{(0,0)}^{\;\;t}&-\tilde{a}_{(1,0)}^{\;\;t}   \\
c_{(0,1)}^{\;\;t}&\tilde{c}_{(1,1)}&-a_{(1,0)}^{\;\;t}&-\tilde{a}_{(0,0)}^{\;\;t} 
\end{array}\right)
 & \hspace{-1em}
		\begin{array}{c}
     p \\ n-p \\[1mm] p \\ n-p  
		\end{array} \\ 
    \mbox{} 
  \end{array} \\[-12pt] , 
\label{ZZso2n}
\end{equation}
where $b_{(1,1)}$, $\tilde{b}_{(1,1)}$, $c_{(1,1)}$ and $\tilde{c}_{(1,1)}$ are antisymmetric matrices.
The dimensions of the four subspaces of $\g=\so_{p}(2n)$ are given by
\[
\begin{array}{ll}
\g_{(0,0)} & p^2+(n-p)^2\\
\g_{(0,1)} & 2p(n-p)\\
\g_{(1,0)} & 2p(n-p)\\ 
\g_{(1,1)} & p(p-1)+(n-p)(n-p-1) .
\end{array}
\]
Once more, the matrices of $\g=\so_{p}(2n)$ do not coincide with the matrices of $G=\so(2n)$ in~\eqref{G}, 
but the dimension of both spaces is the same and equal to $2n^2-n$.

\section{EXAMPLES AND RELATED STRUCTURES} \label{sec:D}%

A very simple example is provided by the Gell-Mann matrices~\cite{Georgi}, usually denoted by
\begin{equation}
\begin{array}{lll}
\lambda_1=\left( \begin{array}{ccc} 0&1&0 \\ 1&0&0 \\ 0&0&0 \end{array}\right)\quad &
\lambda_2=\left( \begin{array}{ccc} 0&-i&0 \\ i&0&0 \\ 0&0&0 \end{array}\right) &
\lambda_3=\left( \begin{array}{ccc} 1&0&0 \\ 0&-1&0 \\ 0&0&0 \end{array}\right) \\[7mm]
\lambda_4=\left( \begin{array}{ccc} 0&0&1 \\ 0&0&0 \\ 1&0&0 \end{array}\right) \quad &
\lambda_5=\left( \begin{array}{ccc} 0&0&-i \\ 0&0&0 \\ i&0&0 \end{array}\right) & \\[7mm]
\lambda_6=\left( \begin{array}{ccc} 0&0&0 \\ 0&0&1 \\ 0&1&0 \end{array}\right) &
\lambda_7=\left( \begin{array}{ccc} 0&0&0 \\ 0&0&-i \\ 0&i&0 \end{array}\right) &
\lambda_8=\frac{1}{\sqrt{3}}\left( \begin{array}{ccc} 1&0&0 \\ 0&1&0 \\ 0&0&-2 \end{array}\right) 
\end{array}
\label{GM}
\end{equation}
Obviously, these 8 elements span the Lie algebra $\ssl(3)$ (in the defining matrix representation). 
Apart from closing under the commutation relations, it is also known that these 8 elements ``almost'' close under anticommutation relations (``almost'' because one also needs the identity matrix to completely close).
They can be seen to completely close under a combination of commutators and anticommutators according to the $\Z_2\times\Z_2$-graded Lie algebra $\g=\ssl_{1,1,1,0}(3)$, i.e.\ matrices of the form~\eqref{ZZsl} with $s=0$ and all other matrix blocks a single element. 
The 8 elements fit into 2-dimensional subspaces according to 
$\g_{(0,0)}=\spn\{\lambda_3,\lambda_8\}$,
$\g_{(1,0)}=\spn\{\lambda_4,\lambda_5\}$,
$\g_{(0,1)}=\spn\{\lambda_1,\lambda_2\}$ and
$\g_{(1,1)}=\spn\{\lambda_6,\lambda_7\}$.
We will not list all the brackets in $\g=\ssl_{1,1,1,0}(3)$, but just the ones coming from anticommutators:
\begin{align*}
&\{\g_{(0,1)},\g_{(1,0)}\}: \quad \{\lambda_1,\lambda_4\}=\lambda_6,\ \{\lambda_1,\lambda_5\}=\lambda_7,\ 
\{\lambda_2,\lambda_4\}=-\lambda_7,\ \{\lambda_2,\lambda_5\}=\lambda_6;\\
&\{\g_{(0,1)},\g_{(1,1)}\}: \quad \{\lambda_1,\lambda_6\}=\lambda_4,\ \{\lambda_1,\lambda_7\}=\lambda_5,\ 
\{\lambda_2,\lambda_6\}=\lambda_5,\ \{\lambda_2,\lambda_7\}=-\lambda_4;\\
&\{\g_{(1,0)},\g_{(1,1)}\}: \quad \{\lambda_4,\lambda_6\}=\lambda_1,\ \{\lambda_4,\lambda_7\}=-\lambda_2,\ 
\{\lambda_5,\lambda_6\}=\lambda_2,\ \{\lambda_5,\lambda_7\}=\lambda_1.
\end{align*}
Although this is a rather straightforward example, the $\Z_2\times\Z_2$-graded structure on~\eqref{GM} combining commutators and anticommutators has not been noticed before.

Another interesting example is provided by an analogue of parafermion operators.
Ordinary parafermion operators satisfy triple commutation relations~\cite{Green,GM}, 
and form a generating set for the Lie algebra $\so(2n+1)$~\cite{KT,Ryan,SV2008}.
Let us consider here the corresponding set of generators from the $\Z_2\times\Z_2$-graded Lie algebra $\so_p(2n+1)$, determined by the last row and last column in~\eqref{ZZso2n+1}. Hence put
\begin{equation}
f_{j}^-= \sqrt{2}(e_{j, 2n+1}-e_{2n+1,n+j}), \quad 
f_{j}^+=\sqrt{2}(e_{2n+1,j}-e_{n+j,2n+1}), \qquad j=1,\ldots , n. 
\end{equation}
As usual, $e_{jk}$ is the notation of a matrix of the relevant size 
with zeros everywhere except a 1 on the intersection of row~$j$ and column~$k$.
In terms of these generators, the subspaces have the following form:
\begin{align*}
& \g_{(0,1)}=\spn \{ f_{k}^{\pm},\; k=1,\ldots,p \} \\
& \g_{(1,0)}=\spn \{f_{k}^{\pm},\; k=p+1,\ldots,n  \} \\
& \g_{(0,0)}=\spn \{ [ f_{k}^\xi,  f_{l}^\eta],\; \xi, \eta =\pm, \; k,l=1,\ldots,p \; \hbox{and} \; k,l=p+1,\ldots,n   \} \\
& \g_{(1,1)}=\spn \{ \{f_{k}^{\xi}, f_{l}^{\eta}\}, \; \xi, \eta =\pm, \;k= 1,\ldots ,p,\; l=p+1,\ldots n \}.
\end{align*}
The set of generating elements $f_i^\pm,\ i=1,\ldots,n$ consists of two sorts of parafermion operators $f_i^\pm,\ i=1,\ldots,p$
and $f_i^\pm,\ i=p+1,\ldots,n$. Each sort of operators is subject to the common defining relations~\cite{SV2008} of parafermion statistics
($j,k,l=1,\ldots,p$ or $j,k,l=p+1,\ldots,n$):
\begin{equation}
[[f_{ j}^{\xi}, f_{ k}^{\eta}], f_{l}^{\epsilon}]=\frac12
(\epsilon -\eta)^2
\delta_{kl} f_{j}^{\xi} -\frac12  (\epsilon -\xi)^2
\delta_{jl}f_{k}^{\eta}, \;\; \xi, \eta, \epsilon =\pm\hbox{ or }\pm 1, \label{PF}
\end{equation}
in terms of nested commutators only.
But the ``relative commutation relations'' between the two sorts of parafermion operators are as follows, and given only in terms of nested anticommutators 
($j=1,\ldots,p,\ k=p+1,\ldots,n, \ l=1,\ldots,n$ or $j=p+1,\ldots,n,\ k=1,\ldots,p, \ l=1,\ldots,n$):
\begin{equation}
\{\{ f_{ j}^{\xi}, f_{ k}^{\eta}\} , f_{l}^{\epsilon}\} =\frac12
(\epsilon -\eta)^2
\delta_{kl} f_{j}^{\xi} +\frac12  (\epsilon -\xi)^2
\delta_{jl}f_{k}^{\eta}, \;\; \xi, \eta, \epsilon =\pm\hbox{ or }\pm 1. \label{PFrel}
\end{equation}
This seems to give a new type of parastatistics, worth studying.
It could describe a model consisting of two sets of parafermionic particles,
where the two sets are not independent of each other but there is a kind of entanglement determined by~\eqref{PFrel}.

In a similar way, one can consider a set of generating relations of the $\Z_2\times\Z_2$-graded Lie algebra $\ssl_{1,q,n-q,0}(n+1)$ from the matrices given in~\eqref{ZZsl}.
Taking
\[
a_j^- = e_{1,j+1},\quad a_j^+=e_{j+1,1}, \qquad j=1,\ldots,n
\]
one finds
\begin{align*}
& \g_{(0,1)}=\spn \{ a_j^-=e_{1,j+1}, \; a_j^+=e_{j+1,1},\ j=1,\ldots ,q \} ,\\
& \g_{(1,0)}=\spn \{ a_j^-=e_{1,j+1}, \; a_j^+=e_{j+1,1},\ j=q+1,\ldots ,n \}, \\
& \g_{(0,0)}=\spn \{ [a_j^+, a_k^-],\ j,k=1,\ldots ,q \;\hbox{and}\; j,k=q+1,\ldots ,n  \} ,\\
& \g_{(1,1)}=\spn \{ \{a_j^-, a_k^+\},\; \{a_j^+, a_k^-\},\ j=1,\ldots ,q \;\hbox{and}\; k=q+1,\ldots ,n  \}.
\end{align*}
The set of generating elements $a_i^\pm,\ i=1,\ldots,n$ consists of two sorts of operators $a_i^\pm,\ i=1,\ldots,q$
and $a_i^\pm,\ i=q+1,\ldots,n$. Each sort of operators is subject to the defining relations of the so-called $A$-statistics~\cite{Jellal}
($j,k,l=1,\ldots,q$ and $j,k,l=q+1,\ldots,n$):
\begin{align}
&[a_j^+,a_k^+]=[a_j^-,a_k^-]=0,\nn \\
&[[a_j^+,a_k^-],a_l^+]=\delta_{jk}a_l^++\delta_{kl}a_j^+, \label{A1R} \\
&[[a_j^+,a_k^-],a_l^-]=-\delta_{jk}a_l^--\delta_{jl}a_k^-,  \nn
\end{align}
purely written in terms of nested commutators. 
The relative relations between the two sorts of operators are however purely in terms of nested anticommutators:
($j=1,\ldots,q,\ k=q+1,\ldots,n, \ l=1,\ldots,n$ and $j=q+1,\ldots,n,\ k=1,\ldots,q, \ l=1,\ldots,n$):
\begin{align}
&\{ a_j^+,a_k^+\}=\{ a_j^-,a_k^-\} =0,\nn \\
&\{ \{ a_j^+,a_k^-\} ,a_l^+\}=\delta_{kl}a_j^+, \label{A1Rel} \\
&\{ \{ a_j^+,a_k^-\},a_l^-\} =\delta_{jl}a_k^-.  \nn
\end{align}

\vskip 5mm
To summarize, we have been able to construct classes of $\Z_2\times\Z_2$-graded Lie algebras corresponding to the classical Lie algebras.
The bracket relations for these new algebras are determined by means of their defining matrices.
For the $\Z_2\times\Z_2$-graded Lie algebra $\ssl_{p,q,r,s}(n+1)$, the construction is not new and coincides with one of the first examples given in the literature.
The $\Z_2\times\Z_2$-graded Lie algebras $\so_p(2n+1)$, $\sp_p(2n)$ and $\so_p(2n)$ are new, 
and determined by their defining matrices given in~\eqref{ZZso2n+1}, \eqref{ZZsp2n} and~\eqref{ZZso2n} respectively.
The structure of these defining matrices is clearly different from their classical counterparts.

Just as the creation and annihilation operator relations of certain types of generalized particle statistics can be associated to generators of classical Lie algebras, the new $\Z_2\times\Z_2$-graded Lie algebras open the way to other types of generalized particle statistics, more specifically to new frameworks for parastatistics.

Let us finally mention that we also attempted to construct a $\Z_2\times\Z_2$-graded Lie algebra analogue of the exceptional Lie algebra $G_2$, using a set of defining matrices of size $7\times 7$ (coming from the embedding $G_2\subset\so(7)$).
Although we were able to identify several candidates for the generating subspaces $\g_{(1,0)}$ and $\g_{(0,1)}$,
using the bracket~\eqref{ZZbracket} we were not able to construct a $\Z_2\times\Z_2$-graded Lie algebra associated to $G_2$.

\section*{ACKNOWLEDGMENTS}

Both authors were supported by the Bulgarian National Science Fund, grant KP-06-N28/6.

\section*{AUTHOR DECLARATIONS}

{\bf Conflict of Interest}

\vskip 0.3cm
\noindent
The authors have no conflicts to disclose.

\vskip 0.5cm
\noindent
{\bf Author Contributions}

\vskip 0.3cm
\noindent
Both authors contributed equally to this work.

\vskip 0.5cm
\noindent
{\bf Data Availability} 

\vskip 0.3cm
\noindent
The data that support the findings of this study are available within the article.

\end{document}